\documentclass[aps,
superscriptaddress,
prl,
amsmath,
nofootinbib,
twocolumn,
amssymb,
10pt
]{revtex4-1}

\usepackage{ulem}
\usepackage{soul}      
\usepackage{color}
\usepackage{txfonts}
\usepackage[colorlinks, linkcolor=blue]{hyperref}
\usepackage{verbatim}

\usepackage{relsize}
\relscale{1.5}

 \usepackage{mathtools}
 \usepackage{wrapfig} 
 \usepackage{mathrsfs}
 \usepackage{amssymb} 
 \usepackage[x11names]{xcolor}
 \usepackage{amsbsy}
 \usepackage{marginnote}
 \usepackage{slashed}
 \usepackage{marvosym}
 \usepackage{pifont}
 \usepackage{mathbbol}
 \usepackage{wrapfig}
 \usepackage{upgreek}
 \usepackage{bbm}
 \usepackage{yfonts}
 \usepackage{xspace}
 \usepackage{lineno}
 \usepackage{tikz}
\usepackage[most]{tcolorbox}
 
 \newcommand{\bcen}{\begin{center}}
 \newcommand{\ecen}{\end{center}}
 \newcommand{\btab}{\begin{tabular}}
 \newcommand{\etab}{\end{tabular}}
 \newcommand{\bdes}{\begin{description}}
 \newcommand{\edes}{\end{description}}

 \newcommand{\beq}{\begin{equation}}
 \newcommand{\eeq}{\end{equation}}
 \newcommand{\bea}{\begin{eqnarray}}
 \newcommand{\eea}{\end{eqnarray}}

 \newcommand{\half}{\frac{1}{2}}
 \newcommand{\bary}{\begin{array}}
 \newcommand{\eary}{\end{array}}
 \newcommand{\benum}{\begin{enumerate}}
 \newcommand{\eenum}{\end{enumerate}}
 \newcommand{\bitem}{\begin{itemize}}
 \newcommand{\eitem}{\end{itemize}}

 \newcommand{\dou}{\partial}

 \newcommand{\D}[1]{\mbox{d}{#1}}

 \newcommand{\eqn}[1] {eqn.~(\ref{#1})}

 \newcommand{\fig}[1]{fig.~\ref{#1}}
 
 \newcommand{\mylabel}[1]{\label{#1}}

\let\chapter\section
\let\section\subsection
\let\subsection\subsubsection

\newcommand{\oibook}[1]{}

\newcommand{\mbbZ}{{\mathbb{Z}}}

\newcommand{\mytitle}{{
$(k,n)$-fractonic Maxwell theory}}

\newcommand{\atrans}[5]{{
\dou_{#2}\dou_{#4}{#1}_{(#3 #5)}
-\dou_{#3}\dou_{#4}{#1}_{(#2 #5)}
-\dou_{#2}\dou_{#5}{#1}_{(#3 #4)}
+\dou_{#3}\dou_{#5}{#1}_{(#2 #4)}
}}

\newcommand{\IISc}{Centre for Condensed Matter Theory, Department of Physics, Indian Institute of Science, Bangalore 56012, India}
\newcommand{\MPIPKS}{Max-Planck-Institut f\"ur Physik komplexer Systeme, Dresden 01187, Germany}

\begin{document}

\title{\mytitle}

\author{Vijay B.~Shenoy}\email{shenoy@iisc.ac.in}\affiliation{\IISc}\affiliation{\MPIPKS}
\author{Roderich Moessner}\email{moessner@pks.mpg.de}\affiliation{\MPIPKS}

\date{\today{}}
\begin{abstract} 
Fractons emerge as charges with
reduced mobility in a new class of gauge theories. Here, we generalise
fractonic theories of $U(1)$ type to what we call $(k,n)$-fractonic
Maxwell theory, which employs  symmetric order-$n$ tensors of
$k$-forms (rank-$k$ antisymmetric tensors) as  ``vector
potentials''. The generalisation has two key manifestations. First,
the objects with mobility restrictions extend beyond simple charges to higher order
multipoles (dipoles, quadrupoles, $\ldots$) all the way to 
$n^\mathrm{th}$-order multipoles. Second, these fractonic charges themselves are characterized by tensorial densities of $(k-1)$-dimensional extended objects. The source-free sector exhibits `photonic' excitations with dispersion $\omega\sim q^n$.
\end{abstract}


\maketitle 




\noindent
{\it Introduction:} Understanding and classifying  the  phases of systems with many interacting microscopic degrees of freedom is gaining renewed attention. It has become clear that, in addition to the traditionally established notions of symmetry, ideas of topology and entanglement are key to obtaining deeper insight and clarity in the classification of phases of many body systems\cite{Wen2017}.  

Apart from the success in classification of short ranged entangled phases of free fermions\cite{Kitaev2009,Ryu2010}, significant progress has been made in the understanding of symmetry protected phases with interactions\cite{Senthil2015}. Long range entangled phases with topological order have also been elaborated\cite{Wen2017}. Systems with topological order provide opportunities for producing a plethora of new physics with exotic excitations and properties that can, for example, be used for various novel applications such as quantum computation\cite{Kitaev2003}. A particularly interesting (and useful) property of topologically ordered systems is that they demonstrate a ground state degeneracy that is determined by the topology of the underlying manifold. The ``topological nature'' of such phases can concretely be appreciated using the example of $\mbbZ_2$ toric code\cite{Kitaev2003} which has a ground state degeneracy of 4 on a torus whether defined using a triangular or square lattice to cover the torus -- only the topology of the torus comes to play in determining the ground state degeneracy. 

Quite fascinatingly, recent research \cite{Chamon2005,Bravyi2011,Castelnovo2012,Castelnovo2012,Haah2011,Yoshida2013,Bravyi2013,Vijay2015,Vijay2016,Williamson2016,Hsieh2017}, broadly following the above cues, has uncovered an  intriguing new kind of phase of matter -- the fracton phase (see \cite{Nandkishore2019} for a review). Fractonic phases derive their name from the peculiar properties of their excitations. These have ``fractional mobility'' in that there are excitations whose motion is  restricted. For instance, the X-cube model, which provides a popular example of fracton physics (see, for example, \cite{Vijay2016}), has point like charge excitations which are immobile, while dipole excitations can move along specific lines. Similar physics, found in all fracton phases, arises not from the energetics, but from the constraint structure encoded in the theory itself that restricts local changes in microscopic degrees of freedom. 
In addition, a typically sub-extensive ground state `entropy' of these systems  is sensitively dependent on the microscopic lattice on which the theory is defined, leading to the definition of a novel concept of ``geometric order''~\cite{Slagle2018}.

The obvious question of where the fracton phases fit in the classification has motivated further work. A natural starting point is to ask for a field theoretical description of such phases that obtains both the excitation physics and the ground state degeneracy. Ref.~\cite{Slagle2017}, starting from the X-cube model, uses the BF-formulation to obtain a field theory that contains the key fractonic physics. Other approaches\cite{You2019} combine the Chern-Simons/BF formulation  with symmetric tensor (see below) gauge formulations to obtain the fracton characterization.  All this clearly points to a  rich panorama of possibilities.

Concurrently with these developments, fractons have appeared in an apparently different context in the study of spin liquids described by symmetric tensor gauge theories\cite{Pretko2017a,Pretko2017b}. Such $U(1)$ symmetric-tensor gauge theories  support a gapless ``photon'' excitation much like Maxwell electrodynamics, which is described by a vector gauge field. Sources(charges) of such a symmetric-tensor gauge theory have a crucial additional character -- an isolated charge of a symmetric-tensor gauge theory is immobile. This arises from the fact that charge conservation in such theories also forces the conservation of dipole moment. As a consequence, motion of an isolated charge, as it changes the dipole moment, is forbidden, endowing the charge with a fracton character.  An isolated dipole (total charge zero) does not suffer such constraints: dipoles are free to move in an unconstrained fashion (to be compared with the dipole excitations in the X-cube model discussed above). Charge conservation is enforced by the continuity equation
\beq\mylabel{eqn:Cont}
\dou_t \rho + \dou_i J^Q_i = 0,
\eeq
where $\rho$ is the charge density, $t$ is  time, $\dou_t$ is time, and $\dou_i$ the spatial derivative along the $i$-th Cartesian coordinate direction, and $J^Q_i$ is the charge current vector. The fractonic nature of charge in symmetric-tensor gauge theories follows from the fact that the charge current vector is itself the divergence of the dipole current tensor $J_{ij}$ , i.~e.,
\beq\mylabel{eqn:Charge}
J^Q_i = \dou_j J_{ji}.
\eeq
Several interesting aspects of such symmetric-tensor gauge theories, including their connection to gravity, have been explored \cite{Pretko2017c,Pretko2017d}. While these developments are interesting and encouraging, future research will have to reveal if such symmetric-tensor gauge theories form a generic framework to describe gapless fracton phases (as noted above, gapped variants starting from symmetric-tensor descriptions are discussed in ref.~\cite{You2019}).

It is compelling to look for generalization of fractonic theories. In the symmetric-tensor gauge theories, charge conservation forces dipole moment conservation. It is then natural to ask if  a hierarchy of such conservation laws can be constructed, where conservation of charge implies conservation of all multipole-moments of the charges, say up to some $n$-th moment (i.e., $2^n$-pole). Physically, this would imply mobility restrictions not only for individual charges, but also on all higher-moment objects such as dipoles, quadrupoles, $\ldots$, $2^n$-pole -- we call this the rank-$n$ fractonic character.  Another desirable generalization of the symmetric-tensor gauge theories is from mobility restricted ``scalar'' charges to ``extended $(k-1)$ dimensional'' objects.

In this paper, we achieve this by formulating what  we call $(k,n)$-fracton theory. These have tensor charge densities of $k-1$-dimensional extended objects with generalized mobility restrictions on all multipoles up to $n$-th order.  This is accomplished by constructing  a new gauge structure -- gauge fields are order $n$ symmetric tensors  of $k$-forms (anti-symmetric tensors) in arbitrary spatial dimension $d$. This leads naturally to a sequence of theories under the heading of $(k,n)$-fractonic Maxwell theory, which for $(k=1,n=1)$ reduces to standard electromagnetism.

We first present a straightforward generalization of the symmetric tensor gauge theories to make $(1,n)$-fracton theories with higher conserved multipole moments of ``scalar'' charges,  followed by the development  of the full 
$(k,n)$-fracton theory. We close with an extended outlook.

\noindent
{\it Notation:} We use a notation that brings out the physical character of our theory that is {\it not} designed for ``relativistic covariance'', with  $x$ denoting a point in $d$-dimensional Cartesian space, and $t$, time. This simple mathematical notation uses symmetric and anti-symmetric tensors. A symmetric tensor of rank-$n$, i., with $n$-indices $i^1,i^2, \ldots,i^n$ is denoted by $S_{(i^1i^2 \ldots i^n)}$ where the $(~)$ used to explicitly indicate the symmetric nature.  On the other hand an anti-symmetric tensor of rank $k$ with indices $i_1,\ldots,i_k$ is denoted by $T_{[i_1,\ldots,i_k]}$ where $[]$ makes the anti-symmetric character of the tensor $T$ explicit. Note that $S_{(i)} = S_{i}$ and $T_{[i]} = T_i$ is to be understood. A collection of $s$ indices $i_1i_2\ldots i_s$ is abbreviated into a composite index $I_s$
\beq
I_s \equiv i_1 i_2 \ldots i_s.
\eeq
Further, $P_s$ of $I_s$ permutes $1 \ldots s$,
\beq
P_s I_s = i_{P_s(1)} \ldots i_{P_s(s)}\ .
\eeq
The sign of the permutation is denoted by $(-1)^{P_s}$.

\noindent
{\bf $(1,n)$-fracton theories:} We  first generalize the symmetric tensor gauge theories of ref.~\cite{Pretko2017a} to rank-$n$ theories, which impose the generalized higher multipole mobility constraints. Consider a gauge field described by $(\phi(x,t),A_{(i^1 i^2 \ldots i^n)}(x,t))$ where the vector potential of the usual Maxwell theory is generalized to a symmetric tensor of rank $n$. We define the electric field tensor (symmetric tensor of rank $n$) as

\beq\mylabel{eqn:EF1n}
E_{(i^1 i^2 \ldots i^n)} = -\dou_{i^1} \dou_{i^2} \ldots \dou_{i^n} \phi - \dou_t A_{(i^1  i^2 \ldots i^n)}.
\eeq
The magnetic field tensor is defined as
\beq\mylabel{eqn:BF1n}
B_{([I^1_2][I^2_2]\ldots[I^n_2])} =  \sum_{\left\{P^r_2 \right\}} \left(\prod_{r=1}^n (-1)^{P^r_2}  \dou_{i^r_{P^r_2(1)}} \right) A_{(i^1_{P^1_{2}(2)}   \ldots i^n_{P^n_{2}(2)})}
\eeq
where $I^r_2 \equiv i^r_1 i^r_2 $, is a composite index and  $P^r_2$ permutes $I^r_2$  in the notation introduced above.
Note that the magnetic field tensor, with a rather unconventional structure, is symmetric in exchange of $I^{r}_2$ with $I^{r'}_2$ composite indices. It is anti-symmetric upon exchange of the two indices within any $I^r_2$.

These fields can be sourced by suitable charges and currents. The charge density $\rho$ and a symmetric $n$-tensor current density $J_{(i^1 i^2 \ldots i^n)}$  couple minimally to the gauge fields. The system is described by a Lagrangian density
\beq\mylabel{eqn:L1n}
L = L_E - L_B + (-1)^n \rho \phi + J_{(i^1 i^2 \ldots i^n)} A_{(i^1 i^2 \ldots i^n)}
\eeq
where repeated indices are contracted ignoring the parentheses (which  simply  remind us of the symmetric nature of the tensors involved). Here $L_E$ and $L_B$ are electric and magnetic energy densities defined by
\beq\mylabel{LE1n}
L_E = \half E_{(i^1 i^2 \ldots i^n)} \epsilon_{((i^1 i^2 \ldots i^n)(j^1 j^2 \ldots j^n))} E_{(j^1 j^2 \ldots j^n)} 
\eeq
where $\epsilon$ is a suitable positive definite dielectric symmetric (in interchange of the set $i$ with $j$) tensor and 
\beq\mylabel{LB1n}
L_B = \half B_{([I^1_2][I^2_2]\ldots[I^n_2])} \kappa_{( (I^1_2][I^2_2]\ldots[I^n_2]) ([J^1_2][J^2_2]\ldots[J^n_2]))} B_{(J^1_2][J^2_2]\ldots[J^n_2])}
\eeq
with $\kappa$ an inverse permeability tensor, again symmetric positive definite. 
The electric and magnetic fields are  invariant under the gauge transformation involving the function $\psi(x,t)$:
\beq\mylabel{eqn:GT1n}
\begin{split}
\phi & \to \phi + \dou_t \psi \\
A_{(i^1 i^2 \ldots i^n)} & \to A_{(i^1 i^2 \ldots i^n)} - \dou_{i^1} \dou_{i^2} \ldots \dou_{i^n} \psi.
\end{split}
\eeq

The gauge invariance of the action (integral of the Lagrangian density) produces the continuity equation
\beq\mylabel{eqn:C1n}
\dou_t \rho + \dou_{i^1} \dou_{i^2} \ldots \dou_{i^n} J_{(i^1 i^2 \ldots i^n)} = 0.
\eeq
This theory thus encodes mobility restrictions not only on charges, but also on any $2^p$-pole (a collection of ``closeby'' charges with leading non vanishing $p$-th multipole moment), $p \le n$, as \eqn{eqn:C1n} implies that
\beq\mylabel{eqn:CL}
\begin{split}
\dou_t \int_V \D{^d x} \, \rho & = 0 \\
\dou_t \int_V \D{^d x} \, x_{i} \, \rho & = 0 \\
\dou_t \int_V \D{^d x} \, x_{i^1} x_{i^2} \, \rho & = 0 \\
& \ldots \\
\dou_t \int_V \D{^d x} \, x_{i^1} x_{i^2}\ldots x_{i^n} 
\, \rho & = 0,
\end{split}
\eeq
all multipoles up to order $n$ are conserved in this theory ($V$ is the $d$-volume of the system). The current $J_{(i^1 \ldots i^n)}$ thus has a natural meaning of the $2^n$-pole current. While charge current is the divergence of the dipole current, the dipole current is in turn that of the quadrupole current, and so on. In this sense, the theory describes rank-$n$ fractonic physics of point charges whose density is $\rho$. 

\noindent
{\bf Theories with extended sources:} The natural next question is if there are generalizations of the theory where the charges are more complex objects. To explore this idea, we first review known Maxwellian theories where  charges are extended objects, a well developed subject in itself (cf.~\cite{KalbRamond1974,Henneaux1986}), casting this theory in our notation. Such gauge theories are described by a gauge field $(\phi_{[i_1 i_2 \ldots i_{k-1}]}(x,t), A_{[i_1 i_2 \ldots i_k]}(x,t))$ where, following our convention, $\phi$ and $A$ are fully anti-symmetric  tensors ($A_{[i_1 i_2 \ldots i_k]}$ is usually called a $k$-form). The electric field is also an anti-symmetric tensor defined by
\beq\mylabel{EFk1}
E_{[i_i i_1 \ldots i_{k}]} = - \frac{1}{(k-1)!} \sum_{P_k} (-1)^{P_k} \dou_{i_{P_k(1)}} \phi_{[i_{P_k(2)} \ldots i_{P_k(k)}]} - \dou_t  A_{[i_1 i_2 \ldots i_k]}.
\eeq
The magnetic field here is obtained as
\beq\mylabel{BFk1}
B_{[i_1 i_2 \ldots i_{k+1}]} = \frac{1}{k!} \sum_{P_{k+1}} (-1)^{P_{k+1}} \dou_{i_{P_{k+1}(1)}} A_{[i_{P_{k+1}(2)} \ldots i_{P_{k+1}(k+1)}]}. 
\eeq
The extended charged sources\footnote{In a discrete lattice, these can be viewed as objects defined using links, plaquettes etc., emanating from a lattice point.\cite{Savit1980}} are described by densities $\rho_{[i_1 i_2 \ldots i_{k-1}]}$ and currents $J_{[i_1 i_2 \ldots i_k]}$. One can now define a Lagrangian density for this theory as
\beq\mylabel{eqn:Lk1}
L = L_E - L_B -  \rho_{[i_1 i_2 \ldots i_{k-1}]} \phi_{[i_1 i_2 \ldots i_{k-1}]} + \frac{1}{k} J_{[i_1 i_2 \ldots i_{k}]} A_{[{i_1 i_2 \ldots i_{k}]}}, 
\eeq
where repeated indices are summed over. The energy densities 
\beq
L_E = \frac{1}{2 k} E_{[i_1 i_2 \ldots i_{k}]} \epsilon_{([i_1 i_2 \ldots i_{k}][i'_1 i'_2 \ldots i'_{k}])}E_{[i'_1 i'_2 \ldots i'_{k}]}
\eeq
and
\beq
L_B = \frac{1}{2(k+1)} B_{[i_1 i_2 \ldots i_{k+1}]} \kappa_{([i_1 i_2 \ldots i_{k+1}][i'_1 i'_2 \ldots i'_{k+1}])}B_{[i'_1 i'_2 \ldots i'_{k+1}]}
\eeq
have suitably defined positive definite dielectric, $\epsilon$, and inverse permeability, $\kappa$, tensors which are both symmetric under the swapping of primed and unprimed indices (with their internal order kept fixed) -- this is the meaning of their subscripts $([i...][i'...])$. 
Again, the electric and magnetic fields are gauge invariant under
\beq
\begin{split}
\phi_{[i_1 i_2 \ldots i_{k-1}]} &= \phi_{[i_1 i_2 \ldots i_{k-1}]} + \dou_t \psi_{[i_1 i_2 \ldots i_{k-1}]} \\
A_{[i_1 i_2 \ldots i_{k}]} &= A_{[i_1 i_2 \ldots i_{k}]} - \frac{1}{(k-1)!} \sum_{P_k} (-1)^{P_k} \dou_{i_{P_k(1)}} \psi_{[i_{P_k(2)} \ldots i_{P_k(k)}]}
\end{split}
\eeq
where $\psi_{[i_1 \ldots i_{k-1}]}(x,t)$ denotes the field that characterizes the gauge transformation. Further, gauge invariance leads to (summing repeated indices):
\beq
\dou_t \rho_{[i_1 i_2 \ldots i_{k-1}]} + \dou_{i_0}J_{[i_0 i_1 \ldots i_{k-1}]} =0.
\eeq
Physically, this implies that the rate of change of the ``density of the extended charge'' is the divergence of its current. Note that this does not yet have any fractonic character; the extended object has no mobility constraints within this theory.  In our nomenclature, this is a $(k,1)$-fracton theory. We  now turn to the question how to endow such extended objects with constrained mobility. In any spatial dimension $d$, we can let $1 \le k \le (d-1)$.

\noindent
{\bf $(k,n)$-fracton theory:} The desideratum of endowing charges with a $(k-1)$-dimensional structure with a rank-$n$ fractonic character is achieved by constructing a suitable gauge structure. We here develop a natural extension of $U(1)$ Maxwell electrodynamics towards this end. The gauge field for $(k,n)$-fracton theory that we propose is of the form 
\beq\mylabel{eqn:GFkn}
(\phi_{([I^1_{k-1}][I^2_{k-1}]\ldots[I^n_{k-1}])}, A_{([I^1_k][I^2_k] \ldots [I^n_k])}).
\eeq
Any field $T_{([I^1_k][I^2_k] \ldots [I^n_k])}$ is fully anti-symmetric under the action of $P_k$ on any $I^r_k$, i.e.,
\beq
T_{([I^1_k][I^2_k] \ldots [P_k I^r_k] \ldots [I^n_k])} = (-1)^{P_k}  T_{([I^1_k][I^2_k] \ldots [I^r_k] \ldots[I^n_k])}.
\eeq
Further, the meaning of the subscript $([I^1_{k}] \ldots [I^n_{k}])$ i.e., $[I^r_k]$s enclosed in $()$ is that it is symmetric under the exchange of any two $I$s. In other words,
\beq
T_{([I^{P_n(1)}_k][I^{P_n(2)}_k]\ldots [I^{P_n(n)}_k])} = T_{([I^1_k][I^2_k]\ldots [I^n_k])}
\eeq
for any permutation $P_n$ of numbers $1 \ldots n$. Note that the notation {\it does not} imply any symmetry in interchanging $i^r_l$ with $i^{r'}_{l'}$, i.~e., two particular indices of $[I^r_k]$ and $[I^{r'}_k]$ ($r \ne r'$).  In other words, the gauge fields that we posit are {``symmetric tensors of anti-symmetric tensors (forms)''}. 

We now define fields in a natural fashion as
\beq \mylabel{eqn:EFkn}
\begin{split}
E_{([I^1_k][I^2_k]\ldots[I^n_k])} = & 
-\frac{1}{[(k-1)!]^n}\sum_{\left\{P^r_k \right\}} \left(\prod_{r=1}^n (-1)^{P^r_k}  \dou_{i^r_{P^r_k(1)}} \right) \phi_{([P^1_k\tilde{I}^1_{k}]\ldots[P^n_k\tilde{I}^n_{k}])}
\\
& - \dou_t A_{([I^1_k][I^2_k] \ldots [I^n_k]))}
\end{split}
\eeq
for the electric field. We have used the notation that 
\beq
P^r_k I^r_k \equiv i^r_{P^r_k(1)}\underbrace{i^r_{P^r_k(2)} \ldots i^r_{P^r_k(k)}}_{\equiv P^r_k \tilde{I}^r_k} 
\eeq
where  $\tilde{I}^r_k \equiv i^r_2 \ldots i^r_k$, i.~e., a composite index with $k-1$ members.
The magnetic field is
\beq\mylabel{eqn:BFkn}
\begin{split}
B_{([I^1_{k+1}]\ldots[I^n_{k+1}])} & = \frac{1}{[k!]^n}\sum_{\left\{P^r_{k+1} \right\}} \left(\prod_{r=1}^n (-1)^{P^r_{k+1}}  \dou_{i^r_{P^r_{k+1}(1)}} \right) A_{([P^1_{k+1}\tilde{I}^1_{k+1}]\ldots[P^n_{k+1}\tilde{I}^n_{k+1}])} \ .
\end{split}
\eeq

The charges of this theory are $\rho_{([I^1_{k-1}] \ldots [I^{n}_{k-1}])}$ and the currents are defined as $J_{([I^1_k]\ldots[I^n_k]))}$ with similar structure as $\phi$ and $A$ respectively. The Lagrangian density is
\beq\mylabel{eqn:Lkn}
\begin{split}
L & = L_E - L_B + (-1)^n \rho_{([I^1_{k-1}]\ldots[I^{n}_{k-1}])} \phi_{([I^1_{k-1}]\ldots[I^{n}_{k-1}])}  \\
 &+ \frac{1}{k^n} J_{([I^1_k]\ldots[I^n_k])} A_{([I^1_k]\ldots[I^n_k])}
\end{split}
\eeq
with
\beq
L_E = \frac{1}{2 k^n} E_\alpha \epsilon_{(\alpha \beta)} E_\beta, \;\; L_B = \frac{1}{2 (k+1)^n} B_\gamma \kappa_{(\gamma \delta)} B_\delta 
\eeq
where $\epsilon$ and $\kappa$ are, again, suitable positive definite dielectric and permeability tensors, $\alpha, \beta$ are indices of the type $([I^1_k] \ldots [I^n_k])$ and $\gamma, \delta$ are indices of the type $([I^1_{k+1}] \ldots [I^n_{k+1}])$ . 
The gauge transformation  described by $\psi_{([I^1_{k-1}][I^2_{k-1}]\ldots[I^n_{k-1}])}$
\beq\mylabel{eqn:GTkn}
\begin{split}
\phi_{([I^1_{k-1}][I^2_{k-1}]\ldots [I^n_{k-1}])} & \to \phi_{([I^1_{k-1}][I^2_{k-1}]\ldots [I^n_{k-1}])} + \dou_t \psi_{([I^1_{k-1}][I^2_{k-1}]\ldots [I^n_{k-1}])} \\
A_{([I^1_k][I^2_k]\ldots[I^n_k])} & \to A_{([I^1_k][I^2_k]\ldots[I^n_k])} \\
& -\frac{1}{[(k-1)!]^n}\sum_{\left\{P^r_k \right\}} \left(\prod_{r=1}^n (-1)^{P^r_k}  \dou_{i^r_{P^r_k(1)}} \right) \psi_{([P^1_k\tilde{I}^1_{k}]\ldots[P^n_k\tilde{I}^n_{k}])}
\end{split}
\eeq
leaves the electric (\eqn{eqn:EFkn}) and magnetic (\eqn{eqn:BFkn}) fields unchanged. The appendix provides the illustration of this general construction for the $(2,2)$-fracton theory. 

From \eqn{eqn:Lkn}, we obtain the equations of motion
\beq\mylabel{eqn:EoM}
\begin{split}
\dou_{i^1_0} \dou_{i^2_0} \ldots \dou_{i^n_0} (\epsilon E)_{([I^1_k][I^2_k]\ldots[I^n_k])} & = \rho_{([I^1_{k-1}] \ldots [I^{n}_{k-1}])}, \\
(-1)^n \dou_{i^1_0} \dou_{i^2_0} \ldots \dou_{i^n_0}  (\kappa B)_{([I^1_{k+1}][I^2_{k+1}]\ldots[I^n_{k+1}])} & \\ 
 = \dou_t (\epsilon E)_{([I^1_k][I^2_k]\ldots[I^n_k])} & + \frac{1}{k^n} J_{([I^1_k][I^2_k]\ldots[I^n_k])}
\end{split} 
\eeq
where $I^r_k \equiv i^r_0 i^r_1 \ldots i^r_{k-1}$; also $i^r_1\ldots i^r_{k-1}$ is identified with $I^r_{k-1}$ (similar convention is used in the second equation, and in \eqn{eqn:Contkn} below).
We further have, from \eqn{eqn:EFkn} and \eqn{eqn:BFkn} that
\beq\mylabel{eqn:Bianchi}
\begin{split}
\frac{1}{[k!]^n}\sum_{\left\{P^r_{k+1} \right\}} \left(\prod_{r=1}^n (-1)^{P^r_{k+1}}  \dou_{i^r_{P^r_{k+1}(1)}} \right) E_{([P^1_{k+1}\tilde{I}^1_{k+1}]\ldots[P^n_{k+1}\tilde{I}^n_{k+1}])} & \\
+ \dou_t B_{([I^1_{k+1}]\ldots[I^n_{k+1}])}  &=0, \\
\sum_{\left\{P^r_{k+2} \right\}} \left(\prod_{r=1}^n (-1)^{P^r_{k+2}}  \dou_{i^r_{P^r_{k+2}(1)}} \right)B_{([P^1_{k+2}\tilde{I}^1_{k+2}]\ldots[P^n_{k+2}\tilde{I}^n_{k+2}])} &= 0
\end{split}
\eeq
The relations \eqn{eqn:EoM} and \eqn{eqn:Bianchi} are the Maxwell equations of our generalized $(k,n)$-fracton theory.
Gauge invariance of the action gives
\beq\mylabel{eqn:Contkn}
\dou_t \rho_{([I^1_{k-1}] \ldots [I^{n}_{k-1}])} +  \dou_{i^1_0} \dou_{i^2_0} \ldots \dou_{i^n_0} J_{([I^1_k][I^2_k]\ldots[I^n_k])} = 0.
\eeq
This achieves the description of the $(k,n)$-fracton density via a tensor $\rho_{([I^1_{k-1}] \ldots [I^{n}_{k-1}])}$. Not only do such ``extended charges'' suffer mobility constraints, but relations similar to \eqn{eqn:CL} also imply mobility constrains on multipoles of order $n$ of such charges. Finally, we find that there are `photonic' excitations in source free $(k,n)$-fracton theory, which disperse as $\omega \sim |q|^n$ ($\omega$ is frequency and $q$ is a wavevector). 

\begin{figure}
    \centering
    \includegraphics[width=\columnwidth]{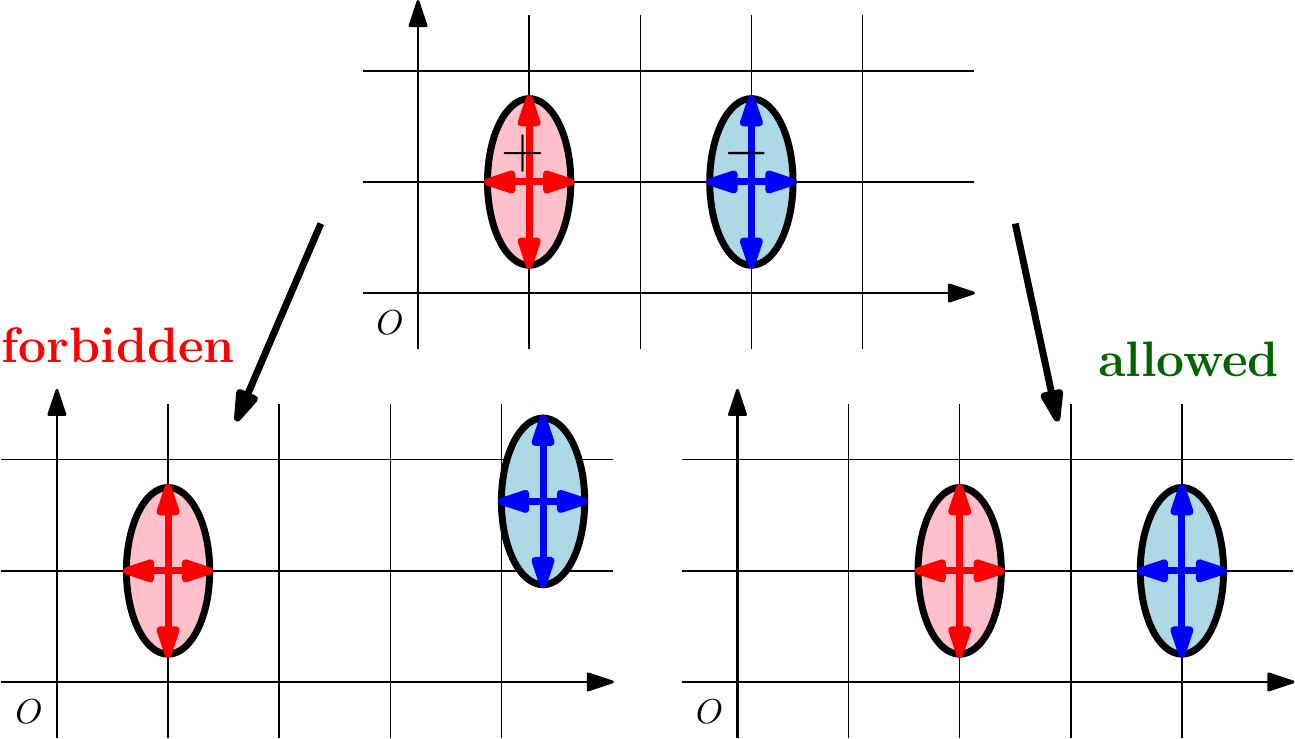}
    \caption{Illustration of tensor charge mobility constraint of (2,2)-fracton theory, see appendix. Charges of this theory are 2-nd rank symmetric tensors, represented by an ellipse indicating principal axes. On the top panel the total charge  vanishes (red is positive charge, blue is negative charge), but the system has a net dipole moment. A rearrangement of the charges which  changes the dipole moment is forbidden (bottom left), while  a dipole-preserving ``rigid'' translations of both charges is allowed (bottom right).}
    \label{fig:frac22}
\end{figure}

An interesting aspect of our construction is the nature of the $(k,n)$-fractonic charges. The character can be better understood by considering the $(2,2)$-fracton example. Here the charges are of the type $\rho_{(ij)}$, i.~e., a symmetric tensor of rank 2. In this theory, individual charges are immobile, but two charges with opposite charge tensors separated in space can move {\it together} as they preserve the net dipole moment. This is pictorially illustrated in \fig{fig:frac22}.

\noindent
{\it Summary and outlook:} In this paper, we have aimed to develop gauge theories that describe extended charges/sources ($k-1$-dimensional) with mobility constraints of rank $n$. We have concluded that a gauge theory with gauge fields that are ``symmetric tensors of anti-symmetric tensors'' provides the desired theory and also fixes the structure of the corresponding $(k,n)$-fracton density.

Our theory, building on \cite{Pretko2017a}, can be viewed as a natural generalization of $U(1)$ Maxwell electromagnetism with freely mobile charges (the $(1,1)$-fracton theory  in our formulation) and constructing a sequence of theories with $(k,n)$-fractons as sources. A number of interesting questions follow immediately from these considerations. 

First, our formulation may not be a unique  generalization of Maxwell electromagnetism, and other routes -- e.g.\ involving symmetric tensors -- leading to yet different kinds of fracton excitations may be worth exploring. In this context, casting our results in the more aesthetical language of differential forms may be worthwhile. 

Second, we have not mandated relativistic invariance for our construction. This is rather natural in condensed matter settings, where such an invariance can effectively emerge, but generically does not, see e.g.\ \cite{MoessnerF2002,Xu2006,Benton2016}. Nonetheless,  identifying possible relativistically covariant generalised fracton theories is surely a worthwhile aim.  

Third, a quantum mechanical extension of generalised fracton theories could unearth novel properties of these theories and their excitations, e.g.\ concerning nature, quantum numbers and statistics of their elementary excitations. This would be embedded in the broader quest for understanding the topological properties of quantum locally constrained models such as quantum dimer models \cite{MoessnerS2001}.

Fourth, what are some microscopic models having generalised fracton theories as effective low-energy description?  
Particularly noteworthy here is the realization that the symmetric tensor gauge theory \cite{Pretko2018} in $d=2$ is dual to elasticity theory (see also \cite{Gromov2017,Gromov2019}). Following this direction, it will be interesting to explore  theories dual to $(k,n)$-fracton theories to look for possible physical realizations. 

Fifth, a related direction is the exploration of the phases of the $(k,n)$-fracton theory and/or their microscopic models of origin. A case in point is, again,  the symmetric tensor gauge theory\cite{Kumar2019,Pretko2019}  in $d=2$ which has provided a fresh perspective into defect driven phase transitions. 

Finally, it is interesting explore the connection of $(k,n)$-fracton theory to some of the discrete fracton phases (as reviewed in the introductory section) with subextensive ground-state degeneracies (see, for example, \cite{Ma2018}), and possible generalization of  lattice gerbe theories\cite{Wegner1971,Savit1980,Lipstein2014,Johnston2014} to lattice tensor-gerbe theories with fractonic physics. A Chern-Simons/BF formulation\cite{You2019} of $(k,n)$-fracton theories could also prove fruitful in obtaining field theories of generalized discrete fracton models.


\noindent
{\it Acknowledgements:} VBS thanks the visitors programme of MPIPKS for hospitality. The authors thank Andr\'es  Schlief and Piotr Sur\'owka for discussions on related work. This work was in part supported by the Deutsche Forschungsgemeinschaft through SFB 1143 (project-id 247310070), and ct.qmat, the  Cluster of Excellence EXC 2147 (project-id 39085490).

\bibliographystyle{apsrev4-1_custom}
\bibliography{fracton}

\appendix

\section{(2,2)-fracton theory}

We illustrate the results for the (2,2)-fracton theory. Gauge fields are
\beq
(\phi_{(ij)} ,A_{([ij][kl])}).
\eeq
The electic field (\eqn{eqn:EFkn})
\beq
E_{([ij][kl])} = -(\dou_i \dou_k \phi_{(jl)} - \dou_j \dou_k \phi_{(il)} - \dou_i \dou_l \phi_{(jk)} + \dou_j \dou_l \phi_{(ik)}) - \dou_t A_{([ij][kl])}.
\eeq
The magnetic field (\eqn{eqn:BFkn})
\beq
\begin{split}
B_{([ijk][lmn])} & =   \dou_i \dou_l A_{([jk][mn])} + \dou_j \dou_l A_{([ki][mn])} +  \dou_k \dou_l A_{([ij][mn])} 
\\
& + \dou_i \dou_m A_{([jk][nl])} + \dou_j \dou_m A_{([ki][nl])} +  \dou_k \dou_m A_{([ij][nl])} \\
& + \dou_i \dou_n A_{([jk][lm])} + \dou_j \dou_n A_{([ki][lm])} +  \dou_k \dou_n A_{([ij][lm])} 
\end{split}
\eeq
The gauge transformation is
\beq
\begin{split}
\phi_{(ij)} & \to \phi_{(ij)} + \dou_t \psi_{(ij)} \\
A_{([ij][kl])} & \to A_{([ij][kl])}  \\
 & - \left(  \dou_i \dou_k \psi_{(jl)} - \dou_j \dou_k \psi_{(il)} - \dou_i \dou_l \psi_{(jk)} + \dou_j \dou_l \psi_{(ik)} \right)
\end{split}
\eeq
Under the gauge transformation 
\beq
\begin{split}
E_{([ij][kl])}  \to &  E_{([ij][kl])} \\
& + \left( - \left[\dou_i \dou_k \dou_t\psi_{(jl)} - \dou_j \dou_k \dou_t\psi_{(il)} - \dou_i \dou_l \dou_t \psi_{(jk)} + \dou_j \dou_l \dou_t\psi_{(ik)}
\right]  \right. \\
& \left. + \dou_t \left[ \dou_i \dou_k \psi_{(jl)} - \dou_j \dou_k \psi_{(il)} - \dou_i \dou_l \psi_{(jk)} + \dou_j \dou_l \psi_{(ik)}\right]
\right) \\
B_{([ijk][lmn])} \to & B_{([ijk][lmn])} \\
- &\left( \dou_i \dou_l \left[ \atrans{\psi}{j}{k}{m}{n}\right] \right. \\
+ & \dou_j \dou_l \left[ \atrans{\psi}{k}{i}{m}{n}\right] \\
+ & \dou_k \dou_l \left[ \atrans{\psi}{i}{j}{m}{n}\right] \\
+ & \dou_i \dou_m \left[ \atrans{\psi}{j}{k}{n}{l}\right] \\
+ & \dou_j \dou_m \left[ \atrans{\psi}{k}{i}{n}{l}\right] \\
+ & \dou_k \dou_m \left[ \atrans{\psi}{i}{j}{n}{l}\right] \\
+ & \dou_i \dou_n \left[ \atrans{\psi}{j}{k}{l}{m}\right] \\
+ & \dou_j \dou_n \left[ \atrans{\psi}{k}{i}{l}{m}\right] \\
+ & \left. \dou_k \dou_n \left[ \atrans{\psi}{i}{j}{l}{m}\right] \,  \right)
\end{split}
\eeq
where the terms in the $\left( \right)$ vanishes ensuring gauge invariant electric and magnetic fields. Gauge invariance of the action will enforce 
\beq
\dou_t \rho_{(ik)} +  \dou_j \dou_l J_{([ji][lk])} = 0.
\eeq

\end{document}